\documentclass[12pt]{iopart}

\usepackage{graphicx}
\usepackage{amssymb}

\pdfoutput=1

\newcommand{\up}{\uparrow}
\newcommand{\down}{\downarrow}
\newcommand{\kF}{k_\mathrm{F}}

\begin{document}

\title[A Measurement Protocol for Entropies of Fermionic Atoms in Optical Lattices]{Thermal vs. Entanglement Entropy: \\
A Measurement Protocol for Fermionic Atoms with a Quantum Gas Microscope}
\author{Hannes Pichler$^{1,2}$, Lars Bonnes$^1$, Andrew J. Daley$^3$, Andreas M. L\"auchli$^1$ and Peter Zoller$^{1,2}$}
\address{$^1$ Institute for Theoretical Physics, University of Innsbruck, A-6020
Innsbruck, Austria}
\address{$^2$ Institute for Quantum Optics and Quantum Information of the Austrian Academy
of Sciences, A-6020 Innsbruck, Austria}
\address{$^3$ Department of Physics and Astronomy, University of Pittsburgh, Pittsburgh,
Pennsylvania 15260, USA}
\ead{hannes.pichler@uibk.ac.at}

\begin{abstract}
We show how to measure the order-two Renyi entropy of many-body states of spinful fermionic atoms in an optical lattice in equilibrium and non-equilibrium situations. 
The proposed scheme relies on the possibility to produce and couple two
copies of the state under investigation, and to measure the occupation
number in a site- and spin-resolved manner, e.g. with a quantum gas microscope.
Such a protocol opens the possibility to measure entanglement and test a number of theoretical predictions, such as area laws and their corrections. As an illustration we discuss the interplay between thermal and entanglement entropy for a one dimensional Fermi-Hubbard model at finite temperature, and its possible measurement in an experiment using the present scheme.
\end{abstract}

\maketitle

%%% INTRODUCTION %%%

\section{Introduction}

Quantum degenerate gases, and ultracold atoms in optical lattices in
particular, provide a unique framework to study quantum many-body physics \cite{Jaksch:2005go,
Bloch:2008gl,Bloch:2012ty,lewenstein2012ultracold}.
This refers first of all to the possibility of controlling many-body
dynamics via external fields, thus allowing one to effectively engineer a wide
class of interesting many-particle Hamiltonians, including those for strongly
correlated systems \cite{Schneider:2008is,Nascimbene:2010en,Jordens:2010kk,Zhang:2012ej}. Furthermore, a plethora of new measurement tools are
available in atomic setups based on probing atoms with laser light,
providing access to physical observables of many-body dynamics,  in a way which is unparalleled in a condensed matter context. An outstanding example is the
recent development of a \textquotedblleft quantum gas
microscope\textquotedblright\ for atoms in optical lattices, which allows
single-atom detection and imaging with resolution of the lattice spacing in
single-shot measurements \cite{Bakr:2009bx,Sherson:2010hg}. Given these unique and novel tools the challenge
is now to identify new atomic measurement protocols that allow access to
new many-particle observables of interest. Below we describe such a
protocol, which allows the direct measurement of the Renyi entropies, quantifying uncertainty due to \textit{thermal fluctuations and due to entanglement},  of (spinful) \textit{fermionic} atoms in
optical lattices for both equilibrium and non-equilibrium situations. The
protocol consists of preparing two identical copies of the many-body systems
in 1D or 2D optical lattices, performing simple single particle operations
that are readily implemented in optical lattices, followed by a read out
with the quantum gas microscope. While in recent work \cite{Daley:2012bd} we have discussed such a
protocol for bosonic atoms, the fermionic case requires rather different
arguments resulting in a different translation table to interpret the
measurement results, although -- quite remarkably -- the basic procedure
parallels the case of bosons.

Direct measurement of thermal and entanglement entropy in atomic gases
brings fundamental concepts, which so far have been discussed exclusively in
a theoretical context, to the laboratory. Examples include the area law
scaling that lies at the heart of the success of matrix product states
methods~\cite{schuch08} and logarithmic corrections in critical systems that
allow us access to properties of the underlying conformal field theory~\cite%
{calabrese04,eisert10a}. Also, the ability to explore higher dimensions is
invaluable as the theoretical understanding of the corrections to the area
law is less advanced and has sparked a lot of recent interest in the theory
community~\cite%
{fradkin06,ryu06,metlitski09,tagliacozzo09,kallin11,metlitski11,humeniuk12,ju12}%
. Other exciting applications include the possibility to detect topological
order~\cite{Kitaev:2006dn,Levin:2006ij} or to monitor the dynamic generation of
entanglement in quantum quenches~\cite{calabrese05,laeuchli08,manmana09,cheneau12} -- 
which are also relevant to questions concerning thermalisation in closed quantum systems~\cite{kinoshita06,hofferberth07,rigol08} -- as
already implemented in cold-atom experiments~\cite{Trotzky:2012iu}. An important
question in this context will be to ask which amount of the entropy is due
to entanglement and what contribution is thermal entropy.
This leads into the question as to whether it is possible to access the quantum entangled regime in current experimental setups, at least for small systems.

To address the latter questions, we use quantum Monte Carlo (QMC) calculations to directly access the finite-temperature Renyi entropies for spinful fermions, described by a Hubbard model in one dimension.
By comparing the finite-$T$ results to ground state density matrix renormalisation group (DMRG) calculations, we study the crossover 
between regimes dominated by quantum entanglement and regimes dominated by thermal entropy for realistic system parameters \cite{Lars}.
These simulation results allow us to give a critical assessment of when such a measurement could
realistically be implemented by addressing the question of limitations of the
measurement protocol. In particular, the exponential scaling of the number of
single measurements with the entropy in the system sets the boundary in term
of temperatures and system sizes, for what will be accessible under realistic
circumstances.

%%% MEASUREMENT PROTOCOL %%%

\section{Measurement Protocol for Renyi Entropies of Fermionic Atoms}\label{Sec:Protocol}

We consider a many-body system represented by fermionic atoms in an optical
lattice. The state of the system at a given time $t$ is described by a
density operator $\rho $. In the following discussion we leave the specific
form of the state open. It can represent a pure state, $\rho =|\psi \rangle
\langle \psi |$, such as the ground state of a Hamiltonian, a thermal state, or
any other (mixed) non-equilibrium state of the fermionic atoms. Our goal is
to develop a protocol to measure Renyi entropies both for the total system
and for subsystems. For the total system the Renyi entropy of order two is
defined by $S_{2}(\rho )=-\log \textrm{Tr}\{\rho ^{2}\}$, and thus is given in
terms of the purity of the density operator $P_{2}(\rho )=\textrm{Tr}\{\rho
^{2}\}$. For a subsystem $\mathcal{R}$ we define a reduced density operator $%
\rho _{\mathcal{R}}=$ Tr$_{\neq \mathcal{R}}\{\rho \}$, and a corresponding
Renyi entropy as $S_{2}(\rho _{\mathcal{R}})=-\log $ Tr$\{\rho
_{\mathcal{R}}^{2}\}$. The knowledge of both $S_2(\rho)$ and $S_{2}(\rho _{\mathcal{R}})$, allows one to quantify the entanglement of the subsystem $\mathcal{R}$ with the rest \cite{Horodecki:2009gb,Mintert:2007gi}. While, as emphasized above, the following discussion is
valid for any quantum state, specific scenarios of experimental interest
include monitoring state purity and entanglement entropies in quench
dynamics as a function of time, or thermal vs.~entanglement entropy in
thermodynamic equilibrium situations.

The Renyi entropy and the purity are nonlinear functionals of the quantum state
and thus not directly observable. However, the purity can be directly
obtained by measuring several copies of the same state \cite{Ekert:2002fp}.
It can be expressed as the expectation value of the \textit{swap operator} $%
V_{2}$ on a system that is prepared in two identical copies in the same
quantum state $\rho $, that is, Tr$\{\rho ^{2}\}=$Tr$\{V_{2}\rho \otimes
\rho \}\equiv \langle V_{2}\rangle $, where the swap operator is defined as $%
V_{2}|\psi _{1}\rangle \otimes |\psi _{2}\rangle =|\psi _{2}\rangle \otimes
|\psi _{1}\rangle $. Similarly, the purity of the reduced density operator
of a subsystem $\mathcal{R}$ is given by the expectation value of the
operator $V_{2}^{\mathcal{R}}$ that swaps the quantum states just in the
part $\mathcal{R}$, that is $\textrm{Tr}\{\rho _{\mathcal{R}}^{2}\}=\textrm{Tr}%
\{V_{2}^{\mathcal{R}}\rho \otimes \rho \}\equiv \langle V_{2}^{\mathcal{R}%
}\rangle $. A measurement of the Renyi entropies $S_{2}(\rho )$ and $S_{2}(\rho _{%
\mathcal{R}})$ of fermionic atoms in optical lattices thus reduces to (i)
the ability to prepare two identical copies of the atomic system, and (ii)
the development of a protocol to measure $\langle V_{2}\rangle $ and $\langle V_{2}^{%
\mathcal{R}}\rangle $ by simple operations and read out in an optical
lattice on the two copies, in a way that can be readily implemented with high
fidelity in present experiments. While the first aspect is primarily an
experimental question, we will focus in the following on the protocol to
realize the measurement of the swap operation (ii) and interpretation of measurement results of
a quantum gas microscope to determine $\langle V_{2}\rangle $ and $\langle V_{2}^{%
\mathcal{R}}\rangle $. We comment on the assumption of
identical copies (i) at the end of this section.

To be more specific, we consider spin-$1/2$ fermions in an optical lattice,
in a setup illustrated for two 2D lattices representing the two copies in
\Fref{fig:setup}. We define a basis of Fock states for the two copies of the system,%
\begin{equation}
|\mathbf{n},\mathbf{m}\rangle \equiv \prod_{(i,\sigma )}{(a_{i,\sigma }^{\dag
})}^{n_{i,\sigma }}\prod_{(j,\sigma )}{(b_{j,\sigma }^{\dag })}^{m_{j,\sigma
}}|\textrm{vac}\rangle ,
\end{equation}%
where $a_{i,\sigma }^{\dagger }$ and $b_{i,\sigma }^{\dagger }$ denote the
creation operations for fermionic atoms on lattice site $i$ and spin $%
\sigma =\pm 1/2$ in the first and second system (copy), respectively, and
where the $\mathbf{n}=\{n_{1,\uparrow },n_{1,\downarrow },n_{2,\uparrow },\dots \}
$ and $\mathbf{m}=\{m_{1,\uparrow },m_{1,\downarrow },m_{2,\uparrow },\dots \}
$ are the occupation numbers of the two systems. The swap operator $V_{2}$ acts on these states as 
$
V_{2}|\mathbf{n},\mathbf{m}\rangle =|\mathbf{m},\mathbf{n}\rangle$,
which interchanges the configurations $\mathbf{n}$ and $\mathbf{m}$. For
bosons this swap operation amounts to the interchange $a_{i,\sigma }^{\dag
}\leftrightarrow b_{i,\sigma }^{\dag }$. Daley \textit{et al}~\cite{Daley:2012bd} have shown
that a measurement of this is readily implemented in an optical lattice by turning on
tunnelling between corresponding lattice sites of the two copies, and reading
out lattice occupation of the first copy (modulo $2$) with the quantum gas
microscope.  However, for fermions applying $V_{2}$ is \textit{not}
equivalent to exchanging $a_{i,\sigma }^{\dag }\leftrightarrow b_{i,\sigma
}^{\dag }$, since there is an ordering problem, i.e. one must keep track of
the fermionic signs. In the following we will present this protocol for fermions. We will present
our results first in the form of an experimental \textit{recipe} in \Sref%
{sec:Protocol}, and give the formal proof in \Sref{sec:Proof} and
\ref{details}. In \Sref{sec:limitations} will analyze limitations and
scaling of errors in these measurements.

\subsection{Experimental Protocol to Measure $\textrm{Tr}\{\protect\rho_{%
\mathcal{R}}^2\}=\textrm{Tr}\{V_2^{\mathcal{R}}\protect\rho\otimes \protect\rho%
\}$}

\label{sec:Protocol}
\begin{figure}[tb]
\centering
\includegraphics[width=0.75\linewidth]{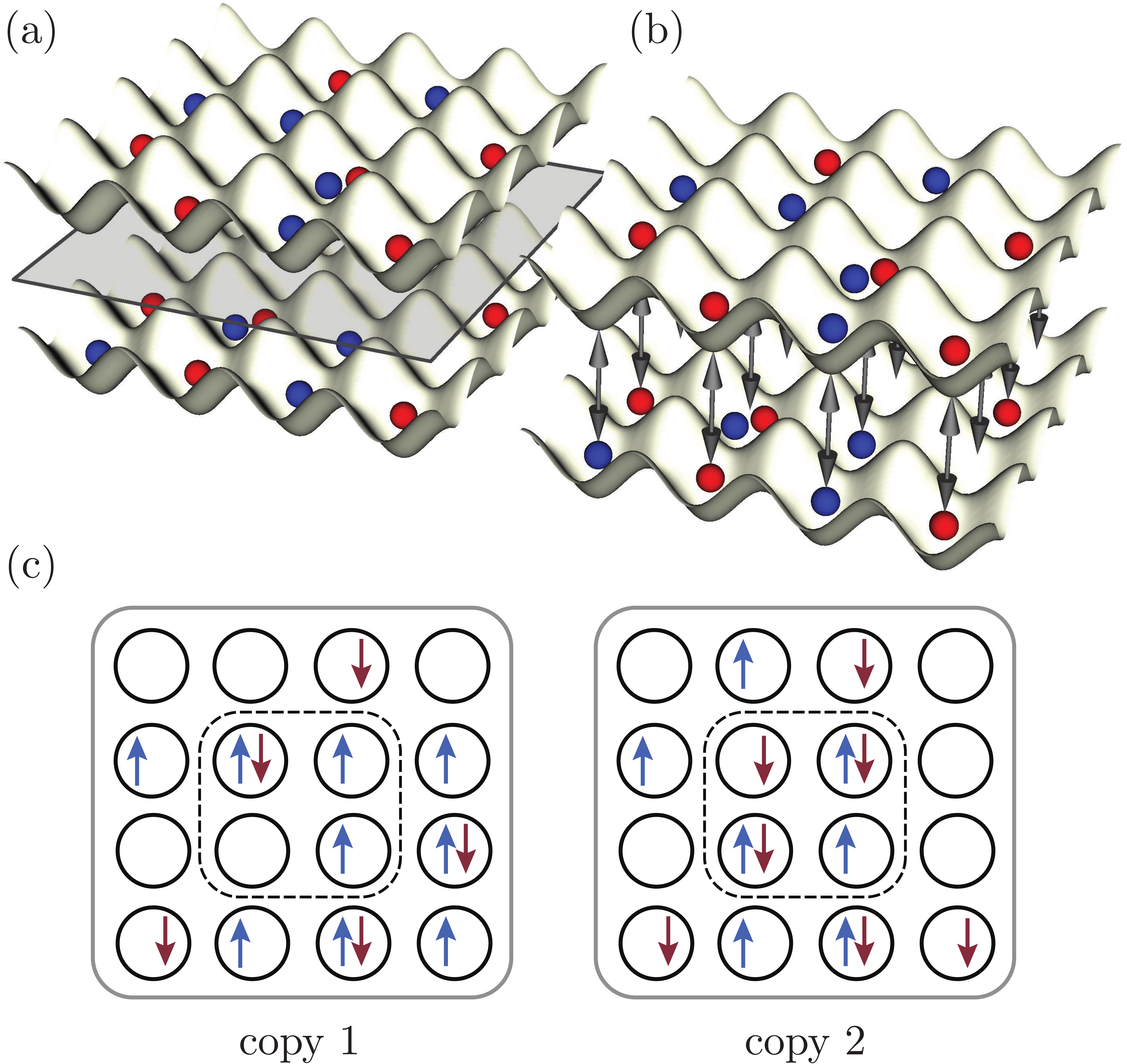}
\caption{(color online) The three steps in the measurement protocol of second
Renyi entropy for spinful fermions in a 2D optical lattice. (a) First, two copies of the
many-body state are produced. (b) Then the barrier between the copies is lowered 
such that atoms can tunnel from each mode to its copy to realize a beam splitter operation between the
copies (see \Eref{eqBS}). (c) Finally, the fermion number is
measured site- and spin-resolved in both copies with a quantum microscope.
Here we show a typical outcome of a single measurement run, where on each
site one finds either no atom, an atom with spin up, an atom with spin down
or both spins on one site. In this example, according to \Tref%
{table1} the measurement outcome for the swap operator on the whole system
is $+1$, while the result for the reduced set of modes that are enclosed by the
dashed line is $-1$. Simultaneously one also obtains measurement results for 
all other subsets. For example the outcome for the swap operator on the spin up modes
is $-1$.}
\label{fig:setup}
\end{figure}

The expectation value of $V_{2}^{\mathcal{R}}$ is obtained by averaging
over a series of single measurements, where each single measurement proceeds
in three steps as illustrated in \Fref{fig:setup}.

(i) Initially two identical copies of the same (non-)
equilibrium state are prepared \cite{Simon:2011ep}. With optical lattices this can be performed in
two parallel 1D tubes or two neighbouring 2D planes that are completely decoupled.
(c.f. \Fref{fig:setup}.a).

(ii) At a given time $t$ the lattice depth within each copy is suddenly
ramped up to freeze the atomic configuration by suppressing tunnelling
(atomic limit). Simultaneously, all interactions (e.g. between different spins) are turned off, e.g. by using magnetic or optical Feshbach resonances \cite{Bolda:2002bd,Theis:2004gk}. Alternatively the protocol can be executed on a timescale where the effects of interactions are negligible. Note that during this step the entanglement and the overall
entropy does not change.

We then lower the barrier between the two copies to
allow tunnelling between each site and its copy with an amplitude $J_{ab}$ for a
fixed time $\tau_{ab}=\pi /(4J_{ab})$ (c.f.~\Fref{fig:setup}.b). This can be accomplished
with the use of an optical superlattice \cite{Folling:2007jg}. With this
operation we implement the beam splitter $U_{2}$ that maps 
\begin{equation}
U_{2}:a_{i,\sigma }\rightarrow \frac{1}{\sqrt{2}}(a_{i,\sigma }+b_{i,\sigma
});b_{i,\sigma }\rightarrow \frac{1}{\sqrt{2}}(b_{i,\sigma }-a_{i,\sigma }),
\label{eqBS}
\end{equation}%
where $a_{i,\sigma }$ denotes the annihilation operator at site $i$ with
spin $\sigma $ and $b_{i,\sigma }$ denotes the annihilation operator in the
corresponding mode in the second copy of the system\footnote{To strictly realise \eref{eqBS} one needs additional phase shifts. They can be obtained by shifting the energies of the two copies relative to each other in an additional step, e.g.\ using a superlattice. However, due to the particle number superselection rule (see \ref{details}) these phase shifts do not affect the final measurement result, and are not necessary for the present protocol.}.

(iii) Finally we measure site- and spin-resolved the occupation numbers using a quantum gas microscope~\cite%
{Bakr:2009bx,Sherson:2010hg}. We denote the measurement results in copy one
by $n_{i,\sigma }^{(1)}$ and in copy two by $n_{i,\sigma }^{(2)}$. In the
case of fermions this number is either zero or one due to the Pauli
principle, but also for bosons it suffices to determine the parity of the
occupation number.

The difference between bosons and fermions consists of how the measured
occupation numbers relate to the measured value of the operators $V_2^{%
\mathcal{R}}$.

\paragraph{Bosons.}

If one finds an even (odd) number of bosons on the modes belonging to $%
\mathcal{R}$ of copy one, then the measurement outcome for $V_{2}^{\mathcal{R%
}}$ is plus (minus) one \cite{Daley:2012bd}. 

\paragraph{Fermions.}

For fermions the measurement outcome of the operator $V_{2}^{\mathcal{R}}$
depends on the total number of fermions in both copies of the modes
belonging to $\mathcal{R}$, which we denote by $N_{\mathrm{tot}}^{\mathcal{R}%
}=\sum_{(i,\sigma )\in \mathcal{R}}n_{i,\sigma }^{(1)}+n_{i,\sigma }^{(2)}$,
and on the number of atoms on copy one of the modes belonging to $\mathcal{R}
$, denoted by $N_{1}^{\mathcal{R}}=\sum_{(i,\sigma )\in \mathcal{R}%
}n_{i,\sigma }^{(1)}$. The corresponding measurement result for $V_{2}^{%
\mathcal{R}}$ can be read off from \tref{table1}. 
\begin{table}[h]
\centering
\begin{tabular}{|c|c|c|c|}\hline
$N_{\rm tot}^{\mathcal{R}}$ & $N_{\rm tot}^{\mathcal{R}}/2$ & $N_{1}^{\mathcal{R}}$ & result for $V_2^{\mathcal{R}}$ \\\hline\hline
 even & even & even & 1\\\hline
 even & even & odd& -1 \\\hline
 even & odd & even & -1 \\\hline
 even & odd & odd & 1 \\\hline
 odd & - & - & 0  \\\hline
\end{tabular}
\caption{Rules determining the measurement outcome for fermions. For a derivation see \Sref{sec:Proof}}\label{table1}
\end{table}
\newline

To determine the expectation value of $V_{2}^{\mathcal{R}}$, and thus the
purity of the corresponding reduced density operator, one has to repeat the
whole measurement procedure and average over the outcomes. Since all the
swap operators for different subsets $\mathcal{R}$ commute $([V_{2}^{%
\mathcal{R}},V_{2}^{\mathcal{R^{\prime }}}]=0)$, it is in principle possible
to measure all of them at once. In fact in each single run one obtains a
measurement result for all possible subsets.

%%% PROOF %%%

\subsection{Justification of the measurement protocol}

\label{sec:Proof} Here we show that the above protocol implements a
measurement of the swap operator and thus the Renyi entropy. We give the
proof for $V_{2}$ and at the end comment on $V_{2}^{\mathcal{R}}$.

As we noted above, the swap operator $V_{2}$ acts on the Fock states
according to $V_{2}|\mathbf{n},\mathbf{m}\rangle =|\mathbf{m,n}\rangle $. In
the bosonic case it is simply given by the operator which interchanges $%
a_{i}^{\dag }\leftrightarrow b_{i}^{\dag }$, since the order of the creation
operators does not matter for bosons. However for fermions applying $V_{2}$
is \textit{not} equivalent to exchanging $a_{i}^{\dag }\leftrightarrow
b_{i}^{\dag }$, since such an operation performs the mapping $|\mathbf{n},%
\mathbf{m}\rangle \rightarrow (-1)^{\sum_{i,j}n_{i}m_{j}}|\mathbf{m},\mathbf{%
n}\rangle $. 

From the Fock-basis one can easily construct the eigenbasis of $V_{2}$. The
eigenspace with eigenvalue $+1$ is spanned by vectors $|\psi _{\mathbf{n},%
\mathbf{m}}^{+}\rangle $ which are of the form $|\psi _{\mathbf{n},\mathbf{n}%
}^{+}\rangle =|\mathbf{n},\mathbf{n}\rangle $, and $|\psi _{\mathbf{n},%
\mathbf{m}}^{+}\rangle =\frac{1}{\sqrt{2}}(|\mathbf{n},\mathbf{m}\rangle +|%
\mathbf{m},\mathbf{n}\rangle )$ for $\mathbf{n}\neq \mathbf{m}$. 
The eigenspace with eigenvalue $-1$ is spanned by the vectors $|\psi _{%
\mathbf{n},\mathbf{m}}^{-}\rangle =\frac{1}{\sqrt{2}}(|\mathbf{n},\mathbf{m}%
\rangle -|\mathbf{m},\mathbf{n}\rangle )$.
Under the beam-splitter operation \Eref{eqBS} the eigenstates of $V_{2}$ transform
into a superposition of states in the occupation number basis according to 
\begin{equation}\label{eqSuperpos}
U_{2}|\psi _{\mathbf{n},\mathbf{m}}^{\pm }\rangle =\sum_{\mathbf{k}}c_{%
\mathbf{k};\mathbf{n},\mathbf{m}}^{\pm }|\mathbf{k},\mathbf{n}+\mathbf{m}-%
\mathbf{k}\rangle .
\end{equation}%
The coefficients $c_{\mathbf{k};\mathbf{n},\mathbf{m}}^{\pm }$ depend on
whether we are considering bosons or fermions. We discuss the two cases
separately.

\textit{Bosons.} The coefficients in \Eref{eqSuperpos} are given by 
\begin{equation}\label{eq_coeff_Boson}
c_{\mathbf{k};\mathbf{n},\mathbf{m}}^{\pm }=\left( 1\pm
(-1)^{\sum_{j}k_{j}}\right) d_{\mathbf{k};\mathbf{n},\mathbf{m}},
\end{equation}%
where $d_{\mathbf{k};\mathbf{n},\mathbf{m}}$ is a numerical factor which is
irrelevant for the following discussion. 
From equations \Eref{eqSuperpos} and \Eref{eq_coeff_Boson} we see that the beam
splitter operation (BS) transforms the symmetric states $|\psi _{\mathbf{n},%
\mathbf{m}}^{+}\rangle $, that is the eigenspace with eigenvalue $+1$ into
the space with an even number of atoms in the first copy, since all
coefficients for states with an odd number of atoms in copy one, $%
\sum_{i}k_{i}$, vanish after the application of the BS. Similarly the
eigenspace with eigenvalue $-1$ is transformed into the space with an odd
number of atoms in the first copy.

\textit{Fermions.} 
We find for fermions 
\begin{equation}
c_{\mathbf{k};\mathbf{n},\mathbf{m}}^{\pm }=\left( 1\pm
(-1)^{\sum_{j}k_{j}+\sum_{i,j}n_{i}m_{j}}\right) e_{\mathbf{k};\mathbf{n},%
\mathbf{m}},
\end{equation}%
where $e_{\mathbf{k};\mathbf{n},\mathbf{m}}$, as for bosons, is a numerical
factor which is irrelevant for our purpose. The main difference to the
bosonic case is the dependence on the parity of the term $%
\sum_{i,j}n_{i}m_{j}$ in the exponent. If this term is even, the situation
is the same as for bosons, and the symmetric states are mapped onto states
with an even number of particles in copy one, while antisymmetric states are
mapped onto states where this number is odd. However if $\sum_{i,j}n_{i}m_{j}
$ is odd, the situation is reversed. 
Unfortunately there is no way of determining the parity of $%
\sum_{i,j}n_{i}m_{j}$ after the beam splitter operation has been applied.
One only has access to the values of $k_{i}$ and $n_{i}+m_{i}$, that
is the number of particles in the modes of copy one after the beam splitter,
and the total number of atoms in both copies of each mode (which is the same
before and after the beam splitter operation). However, we will show in the
following that this is sufficient information to proceed. The basic idea is the
following.

(i) First consider only the eigenstates $|\psi _{\mathbf{n},\mathbf{m}}^{\pm
}\rangle $ with $\sum_{i}n_{i}=\sum_{i}m_{i}\equiv N$, that is with an equal number of atoms in the two systems (before the beam splitter operation is
applied). For these eigenstates the parity of $\sum_{i,j}n_{i}m_{j}=N^{2}$
can be determined from the knowledge of the total number of atoms $N_{%
\mathrm{tot}}=2N$, which can be accessed from the occupation number
measurement in the final step of our protocol. If $N$ is even, then $N^{2}$
is even as well, and as in the bosonic case, the (anti-) symmetric states
are mapped onto the space with (odd) even number of particles in copy one.
On the other hand, if $N$ is odd, then also $N^{2}$ is odd, and the
situation is reversed, the (anti-) symmetric states are mapped onto the
space with (even) odd number of particles in copy one. This is reflected in
the rules presented in \tref{table1}.

(ii) Second, consider only the eigenstates $|\psi _{\mathbf{n},\mathbf{m}%
}^{\pm }\rangle $ with $\sum_{i}n_{i}\neq \sum_{i}m_{i}$. Then the total
number of fermions $N_{\mathrm{tot}}$ is not enough to determine the parity
of $\sum_{i,j}n_{i}m_{j}$. Thus, when assigning a measurement outcome
according to \tref{table1}, there are some pairs $(\mathbf{n},\mathbf{m}%
)$ where one incorrectly assigns a measurement result of $+1$ to the state $%
|\psi _{\mathbf{n},\mathbf{m}}^{-}\rangle $ and a value of $-1$ to the
corresponding state $|\psi _{\mathbf{n},\mathbf{m}}^{+}\rangle $. However,
for product states constrained by a particle number superselection rule \cite{Schuch:2004kc} it is readily shown (see \ref{details})
that the probability of finding the system in the symmetric state $|\psi _{%
\mathbf{n},\mathbf{m}}^{+}\rangle $ is the same as the probability of
finding it in the corresponding antisymmetric state $|\psi _{\mathbf{n},%
\mathbf{m}}^{-}\rangle $, if $\sum_{i}n_{i}\neq \sum_{i}m_{i}$. Therefore
these instances average to zero and the error is irrelevant for the average
value. Furthermore, if the total number of atoms $N_{\mathrm{tot}}$ is odd,
then one can be sure that $\sum_{i}n_{i}\neq \sum_{i}m_{i}$, and assign a
measurement value of zero right away, as suggested in \tref{table1}. A
more formal and detailed proof of these points can be found in \ref%
{details}. In contrast to the bosonic case \cite{Daley:2012bd}, the above arguments can not be generalised in a straightforward way to higher order Renyi entropies for fermions.

If one is interested in the purity of the reduced density operator of a
subset of modes $\mathcal{R}$, that is in the measurement of $V_{2}^{%
\mathcal{R}}$, the discussion is completely analogous. 
Note that, since the the beam splitter is unitary and the beam splitting
operations for different modes commute, it is irrelevant whether or not one
performs the beam splitter operation also on modes not belonging to $%
\mathcal{R}$. Thus in practice one can perform always the full beam splitter 
$U_{2}$ on each pair of modes, and determine the number of atoms and the
corresponding measurement result for $V_{2}^{\mathcal{R}}$ in each
subdivision $\mathcal{R}$ simultaneously.

%%% LIMITATIONS %%%

\subsection{Limitations and Effect of Errors}
\label{sec:limitations} 
The main limitation of the proposed measurement
scheme is that the number of single measurements necessary to determine the
entropy with a certain statistical accuracy can become large. This number can be estimated as follows. Each single measurement gives, according to \tref{table1} either $\pm 1$
or $0$. Their mean value determines $\langle V_{2}^{\mathcal{R}}\rangle =%
\textrm{Tr}\{\rho _{\mathcal{R}}^{2}\}$. The variance of the measurement
outcomes can be expressed as $(\Delta V_{2}^{\mathcal{R}})^{2}=\langle P_{%
\mathrm{even}}\rangle -\langle V_{2}^{\mathcal{R}}\rangle ^{2}$, where $P_{%
\mathrm{even}}$ is the projector on the sector with even total number of
particles in the two copies. 
It is easily shown for product states of two
copies, both constrained by a particle number superselection rule, that $1/2\leq \langle P_{\mathrm{even}}\rangle \leq 1$. To determine
the expectation value of $V_{2}^{\mathcal{R}}$ and thus the purity with a
certain statistical accuracy $\sigma _{V}$, one needs a number of
measurements $\#_{V}$ that is given by $\#_{V}\sigma _{V}^{2}=\langle P_{%
\mathrm{even}}\rangle /\langle V_{2}^{\mathcal{R}}\rangle ^{2}-1$. For
highly mixed states this number diverges as $\#_{V}\sigma _{V}^{2}\sim 1/%
\textrm{Tr}\{\rho ^{2}\}$. This results in an exponential scaling of the
required number of measurements with the Renyi entropy. To determine the
entropy with a relative statistical uncertainty of $\sigma _{S}$ one needs $\#_{S}$
measurements with $\#_{S}\sigma _{S}^{2}=\frac{1}{S_{2}^{2}}(\langle P_{%
\mathrm{even}}\rangle e^{2S_{2}}-1)\sim e^{2S_{2}-2\log S_{2}}$. We note that
all known schemes to measure entropy based on multiple copies 
\cite{Ekert:2002fp,Daley:2012bd,Abanin:2012bj} suffer from this limitation.

Since the (parity of the) number of atoms on each site enters crucially in
the measurement result, errors in the measurement of this numbers are a
major error source in an experiment. Their effect is most easily outlined in
the measurement scheme for bosons, but the discussion can be carried out
analogously for fermions. In an ideal experimental implementation the system
state determines the probability $p_{\pm }$ of finding an eigenvalue plus or
minus one, such that $\langle V_{2}\rangle =p_{+}-p_{-}$. Suppose that with
a probability $\epsilon $ the quantum gas microscope incorrectly measures an
even (odd) number of particles on a certain site, when there is actually an
odd (even) number of atoms. Assuming that such errors occur in an uncorrelated fashion and
with the same probability on the $M$ sites that are measured, the
experimentally determined expectation value $\langle V_{2}\rangle _{\epsilon
}$ is reduced compared to the actual one by $\langle V_{2}\rangle _{\epsilon
}=\langle V_{2}\rangle (1-2\epsilon )^{M}$. Thus the measured entropy $%
S_{2,\epsilon }(\rho )$ is just the sum of the actual entropy of the system $%
S_{2}(\rho )$ and a contribution from 
the quantum gas microscope $S_{\textrm{microscope}}=-M\log (1-2\epsilon )$,
where the contribution of the quantum gas microscope is extensive in the
size of the measured (sub)system. Thus, the measured purity (entropy) is
always smaller (larger) than the actual one.

A general assumption \cite{vanEnk:2009eo} underlying the protocol is the
preparation of two perfect copies $\rho \otimes \rho $. Even though our protocol does not provide a direct means to check this assumption, it can strengthen it a
posteriori, e.g. if the two copies are close to a pure state. Then the
measured expectation value is one if the state in both copies is the same.
Also, one can relax the assumption of having completely uncorrelated copies and allow for classical correlations between them, such that the total density operator is of the form $\rho_{\rm tot}=\sum_i p_i \rho_i\otimes\rho_i$, and still obtain useful information from
the described measurement protocol as outlined in \ref%
{mixtures_of_copies}. On the other hand, one might be interested in the
overlap of two states that are different from the outset. By preparing the
two states in the form $\rho _{1}\otimes \rho _{2}$, this protocol gives
access to $\textrm{Tr}\{\rho _{1}\rho _{2}\}=\textrm{Tr}\{V_{2}\rho _{1}\otimes
\rho _{2}\}$.

%% NUMERICAL RESULTS %%%
\section{Renyi Entropies of a
Fermionic Hubbard Chain}

The protocol to measure Renyi entropies for fermionic systems described in the previous section provides a novel tool to access in experiments fundamental properties of many-body systems related to entanglement and thermal entropy. 
One of the important results quantum information theory has brought to the field of
quantum many-body systems is the finding that ground states of local
Hamiltonians typically exhibit an entropic area law $S \sim \alpha \partial A$, where $\partial A$
denotes the perimeter of the boundary delimiting the two complementary subregions $A$ and $B$~\cite{eisert10}.
In the particular case of one-dimensional critical systems admitting a conformal field
theory (CFT) description, the area-law picks up an (additive) logarithmic correction, whose prefactor
solely depends on the central charge $c$ of the CFT. 
These are very important theoretical results underlying the success of matrix- and tensor network based 
numerical and conceptional methods, and have furthermore deep connections to quantum field theory, 
string theory and black hole physics. It is therefore highly desirable to test these theoretical predictions 
in actual experiments, enabled through the experimental protocols introduced in previous~\cite{Ekert:2002fp,Daley:2012bd,Abanin:2012bj} and the present work.
 
In the present section we illustrate this for the example of the Fermi-Hubbard model in one dimension, as it is the simplest model of interacting fermions that can be realised with cold fermionic atoms in an optical lattice. In terms of  creation ($a_{i,\sigma}^\dagger$), annihilation ($a_{i,\sigma}$) and counting operators ($n_{i,\sigma}=a_{i,\sigma}^\dagger a_{i,\sigma}$) the Fermi-Hubbard Hamiltonian for a 1D lattice with $L$ sites is given by 
\begin{equation}
 H=-t_F \sum_{\sigma=\uparrow,\downarrow} \sum_{i=1}^{L-1} \left(
 a^\dagger_{i,\sigma} a_{i+1,\sigma} + \mathrm{h.c.} \right) + U \sum_{i=1}^L
 n_{i,\uparrow} n_{i,\downarrow},
 \end{equation}
 where $t_F$ denotes the hopping amplitude between neighbouring sites and $U$ the onsite interaction energy.
 The model is exactly solvable via the Bethe ansatz~%
 \cite{lieb68,essler05} and the phase diagram exhibits a metallic two-channel Luttinger
 liquid ground state at generic fillings for all interactions $U\geq 0$ ~\cite{giamarchi04,essler05}. At half filling and $U>0$ the charge degrees of freedom are gapped and a Mott insulator appears.
 
Here, we use a generalised directed
loop algorithm within the stochastic series expansion (SSE) framework~\cite%
{sandvik99b,syljuasen02,alet05} to access the thermal Renyi entropies following a measurement scheme based on the dynamic update of the world line topologies presented in Ref.~\cite{humeniuk12}.
Large blocks are built up consecutively using the \textit{increment trick}~\cite{hastings10} that allows for an efficient update in replica space.

In the following we quantify the ground-state entanglement as well as the thermal Renyi entropy that one would obtain with the proposed protocol for realistic experimental system sizes. We show to what extent quantum entanglement can be accessed through a measurement of $S_2$ for  system sizes and temperatures available in an experiment. A systematic study of the crossover between entanglement and thermal entropy is presented in \cite{Lars}.

\subsection{Zero Temperature}
 \begin{figure}[tb]
 \centering
 \includegraphics[width=0.75\linewidth]{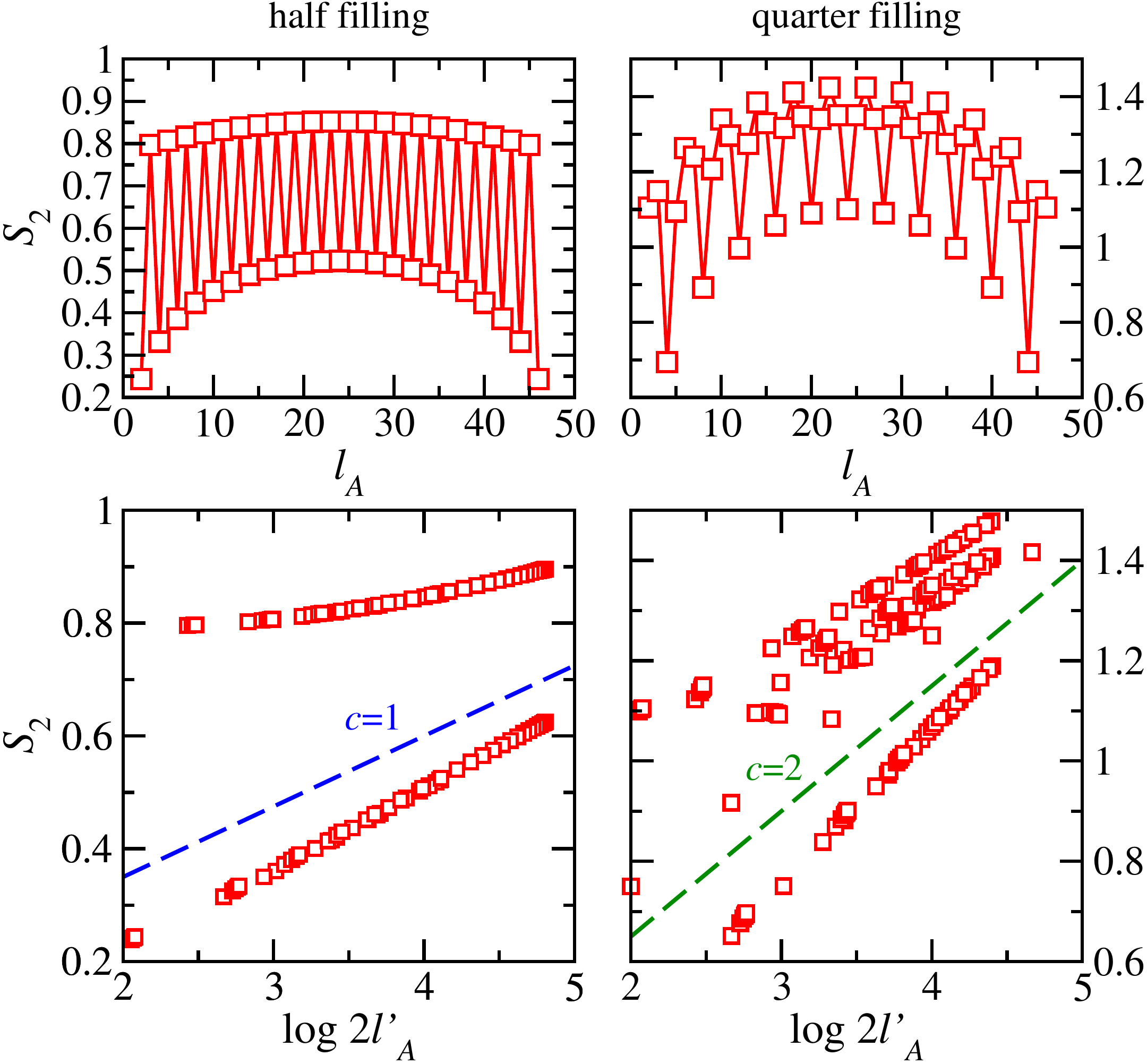}
 \caption{The upper left (right) panel shows the second Renyi entropy $S_2$ for the half-filled (quarter-filled) Hubbard model as a function of the chord distance $l'_A$ for $L=48$ and $U/t_F=8$ obtained using DMRG.
 The lower panel displays the same entropy data as a function of $\log l'_A$, where $l'_A$ is the chord length of block $A$. Here, we show DMRG data for systems up to $L=96$.
 The slope of $S_2$ is given by $c/8$ (see \Eref{eq:RenyiOsc}) and the blue and green lines are guides to the eye  corresponding to $c=1$ and $c=2$ respectively.
 }
 \label{fig:EntropyScaling}
\end{figure}
 Let us start the by looking at the Hubbard chain at zero temperature. 
\Fref{fig:EntropyScaling}  shows the $n=2$ Renyi profiles for bipartitions $A \cup B$ with block sizes $l_A$ obtained using DMRG for chains with $L=48$ sites, both for half ($n_\up = n_\down = 1/2$) and quarter filling ($n_\up = n_\down = 1/4$). A prominent feature of the the Renyi profiles for finite system sizes is that they exhibit characteristic oscillation associated with the Fermi-momentum $\kF$, giving rise to two (four) branches at half (quarter) filling. Further, one clearly identifies the envelope carrying the logarithmic corrections to the area law.

These features are well understood via the underlying conformal field theory
 \cite{calabrese09,calabrese10, calabrese10a,fagotti11}, from which the Renyi profiles are obtained to be
 \begin{equation}
 S_n(A) = \frac{c}{12} \left(1+\frac{1}{n} \right) \log 
 2 l^{\prime}_A + S_n^\mathrm{corr}(l^{\prime}_A) + \mathrm{const.},
 \label{eq:RenyiOsc}
 \end{equation}
where $l^{\prime}_A = L/\pi \sin(\pi l_A/L)$ is the chord distance. 
The first term is the leading contribution and describes the logarithmic
increase of the entropy with block size. It is directly proportional to the
central charge $c$ and distinguishes between the metal ($c=2$) and the
Mott insulating state ($c=1$) having two respectively one gapless channels. 
By measuring the Renyi profiles with the previously introduced protocol it is in principle possible to determine this pre-factor and to extract the central charge $c$ in an experiment.
This is conveniently done by plotting $S_2$ as a function of $\log(2l'_A)$, as shown in \Fref{fig:EntropyScaling}.
This way, $S_2$ approaches a straight line with slope $c/12(1+1/n)$ for large block and system sizes.
One can see from \Fref{fig:EntropyScaling} that the data for the half filled Hubbard Model at $U/t_F=8$ is consistent with a central charge of $c=1$, as the two branches of the Renyi entropy approach the corresponding asymptotic line from above and below whereas the quarter filled chain exhibits $c=2$.

\subsection{Finite Temperature and Trap}
\begin{figure}[tb]
\centering
\includegraphics[width=0.8\linewidth]{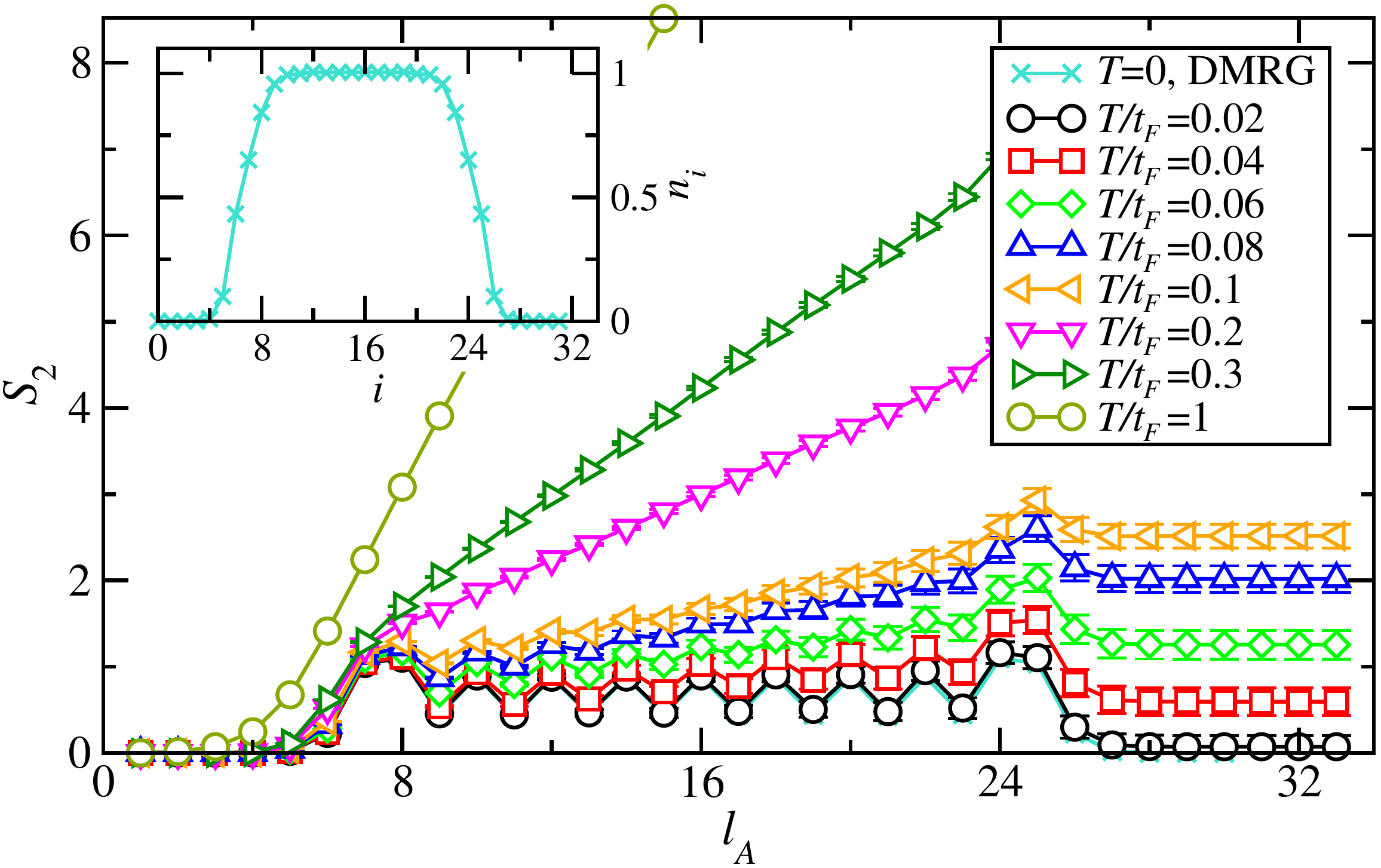}
\caption{Renyi profile for a Hubbard chain in the presence of a trapping
potential with $V(i)=(i-L/2)^2/20 - 6$ and $U/t_F=6$. The green crosses correspond the ground-state of the
system obtained from DMRG with 18 particles whereas the finite-temperature
results are obtained from QMC using grand-canonical simulations with an
average particle number coinciding to the $T=0$ DMRG calculations. A density
profile for different temperatures is shown in the inset. 
}
\label{fig:Trap}
\end{figure}

A measurement of the Renyi Profile as proposed in \Sref{Sec:Protocol} will necessarily be performed at finite temperature.
Then, the Renyi entropy will not only pick up quantum entanglement but also thermal contributions to the entropy. 
Since the thermal entropy is an extensive quantity, one expects a crossover from the area-law to a volume law.
This is shown in \Fref{fig:Trap}, where we plot 
 finite-temperature QMC results for $U/t_F=6$. 
 To make a closer connection to the experimental situation, we include a
harmonic trapping potential, $H_\mathrm{trap} = \sum_i V_i n_i$, with $V(i) = (i-L/2)^2/20-6$. Filling the trap with 18 particles, at zero temperature a
Mott insulator extends over about eight lattices sites and the density drops
to zero in the wings quite rapidly (see the density profile in the inset of
\Fref{fig:Trap}). In the centre of the trap at $T=0$, where the Mott phase proliferates, the Renyi profile shows the characteristic $2k_%
\mathrm{F}$ oscillations. The metallic wings that appear due to the presence of the harmonic trap, are also clearly visible in the Renyi profile in \Fref{fig:Trap}.

When the temperature is increased above the finite size gap,
the entropy $S_2$ picks up a linear contribution (see \Fref{fig:Trap}).
Thus, the $S_2$ profile is no longer symmetric with respect to the centre of the system and the purity of the whole system decreases as $S_2(L)$ increases. In addition the amplitude of the parity induced oscillations is suppressed as the temperature is increased.
One can see that for temperature $T/t_F \gtrsim 0.1$, the entropy is already dominated by a linear increase of the thermal entropy and $S_2$ looses its parity effects.

The ability to measure Renyi profiles also allows access to other quantities that provide further insight into the entanglement and correlation properties of the system.
For instance, the mutual information between two (possibly disjoint) blocks $A$ and $B$ and is given in terms of the Renyi entropies as
\begin{equation}
 I_2(A|B) = S_2(A) + S_2(B) - S_2(A \cup B).
\end{equation}
This measure is particularly useful when dealing with mixed states as it does not pick up an extensive contribution from the thermal entropy -- it obeys an area law~\cite{srednicki93,amico08,eisert10a} even at finite-$T$ -- but is sensitive towards correlations between the two subblocks~\cite{wolf08}.
For a detailed discussion of the mutual information in the Hubbard chain we refer to Ref.~\cite{Lars}.

\subsection{Limitations}
\begin{figure}[tb]
\centering
\includegraphics[width=0.8\linewidth]{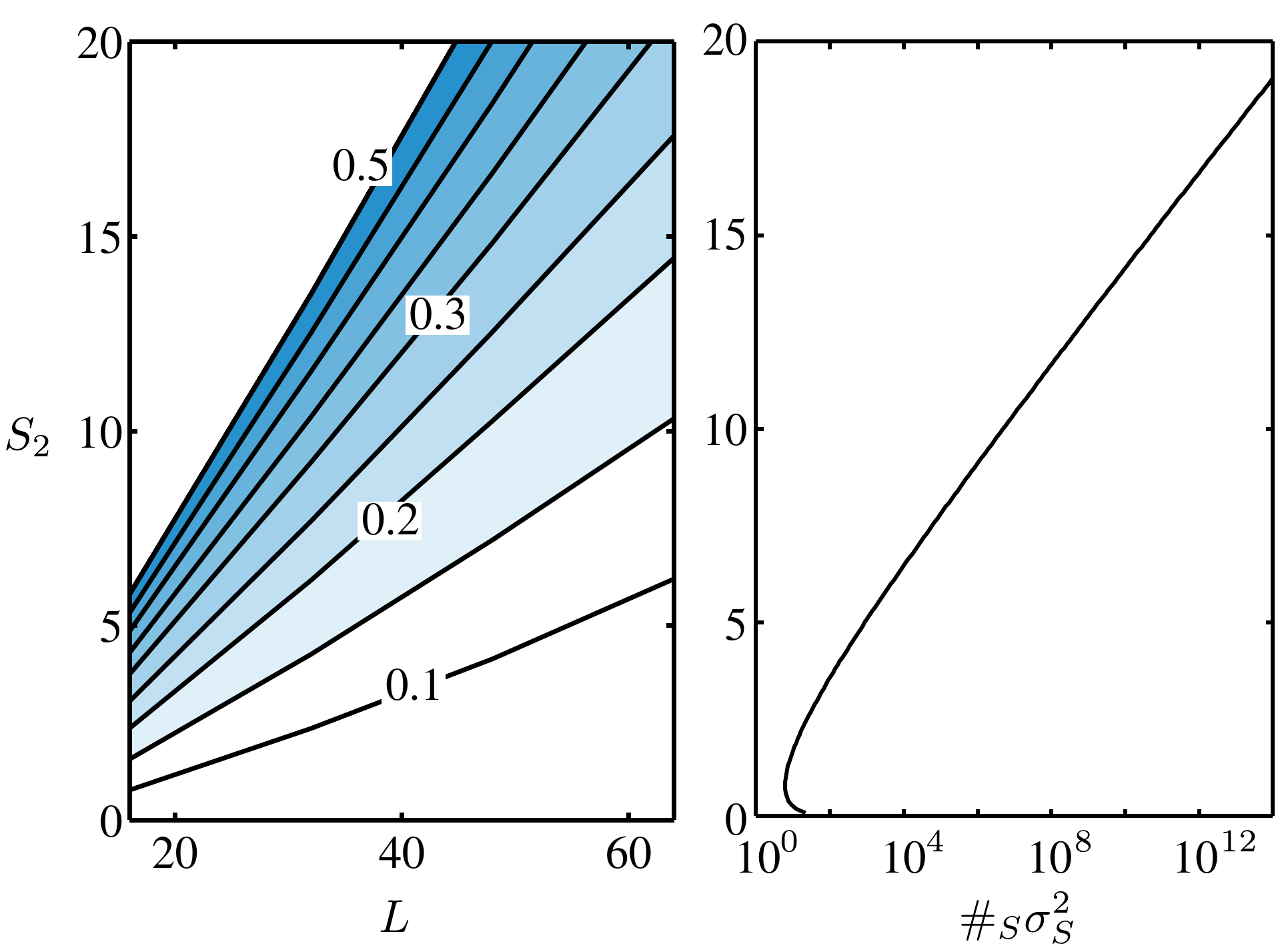}
\caption{(left) Renyi entropy of the whole chain, $S_2(L)$ at quarter filling and $U/t_F=4$, for different system sizes $L$ at different temperatures $T$. The contour lines denote constant temperature $T/t_F=0.1$, 0.15, ... 0.5.(right) Number of measurements, $\#_S$,  required for a measurement of $S_2$ with a statistical relative error $\sigma_S$, using our protocol.}
\label{fig:Numbers2}
\end{figure}
As pointed out in \Sref{sec:limitations}, the ability to determine these entropy profiles 
using the protocol proposed in this work is fundamentally limited by an 
exponential growth with entropy of the number of single measurements required to obtain a 
certain statistical accuracy $\sigma_S$ (see \Fref{fig:Numbers2}). To give numbers, for example $10^5$ 
single measurements are needed to determine Renyi entropies up to $S_2\sim 5$ with a relative statistical error of $\sigma_S\sim0.1$. 
For one dimensional systems, at temperatures below the finite size gap this is not a sever restriction as the entropy obeys a area law with at most logarithmic corrections. For example in the previous subsection we showed that in the 1D Hubbard model, at temperatures 
below the finite size gap the entropies are typically of the order one. This values can be resolved with just $\sim10/\sigma_S^2$ single measurements. 
Above the finite size gap however the number of measurements sets the limit in terms of temperatures and system sizes, as the entropy becomes extensive in the system size and increases with temperature. 

To quantify this, \Fref{fig:Numbers2} shows the Renyi entropy $S_2(L)$ of a full (homogenous) Hubbard chain at quarter filling and $U/t_F=4$ as a function of system size $L$ (open boundary conditions) and temperature $T$, as well as the number of measurements, $\#_S$,  required to resolve a certain value of $S_2$ with a statistical relative error, $\sigma_S$, using the above proposed protocol.  This clearly shows that experimentally relevant situations, e.g. $T/t_F\sim0.2$, $L\sim 30$ can be explored with a moderate number $\#_S\sigma_S^2\lesssim 10^3$ of single measurements.

\section{Discussion}
To summarise, we have presented a method to determine the order two Renyi Entropy for
bosons and fermions in an optical lattice. The scheme is based on the
possibility of preparing two identical copies of a quantum state in an
optical lattice, coupling the two copies via a superlattice, and site-(and
spin-) resolved measurement of the occupation number (modulo two). The
combination of these tools, which are available in current experiments, allows
to directly determine the entropy of a many-body quantum state, and may be
seen as a thermometer for states near the absolute ground state. On the
other hand, for pure states this opens the possibility to study entanglement
as quantified by the entropy of a subsystem. Possible applications include
test area laws for the scaling of entanglement entropy in ground states, and
monitoring entanglement in time-dependent systems.
Using QMC-techniques we analyzed the possibility to test area laws and their 
corrections in a finite size system at finite temperatures typically present in 
current experiments.

\ack
We thank I. Bloch, M. Greiner, C. Gro\ss, and M. Zwierlein, for helpful and motivating discussions.
Work in Innsbruck is supported in part by the EU project AQUTE, the Austrian Science Fund (FWF) through SFB  FOQUS and the Austrian Ministry of Science BMWF as part of the UniInfrastrukturprogramm of the Research Platform Scientific Computing at the University of Innsbruck. Work in Pittsburgh is supported by a grant from the US Army Research
Office with funding from the DARPA OLE program and by NSF Grant PHY-1148957.

\appendix

\section{Details on the Proof of the Measurement Protocol}

\label{details} Here we elaborate more formally on some points of the proof
presented in \Sref{sec:Proof} for fermions. First, note that a pure state (in one copy) with $N$
fermions (constrained by the particle number superselection rule) always has the form: 
\begin{equation}
|\psi \rangle=\sum_{\tiny{\left.\begin{array}{c}\textbf{n} \\ \sum_i n_i=N\end{array}\right. }} \psi_{%
\mathbf{n}}|\mathbf{n} \rangle,  \label{eq8}
\end{equation}
where $N$ is the total number of atoms. A general state is a mixture of such
states (with possibly different total number of fermions): 
\begin{equation}
\rho=\sum_{{\tiny{N}}}\sum_{\tiny{\left.\begin{array}{c} \textbf{n},\textbf{n'}
\\ \sum_i n_i=N \\ \sum_i n_i'=N\end{array}\right.}}\rho_{\mathbf{n},\mathbf{n'}}^{(N)}|\mathbf{%
n} \rangle\langle \mathbf{n'} |.  \label{eq9}
\end{equation}
Thus, a general \textit{product} state in the two systems has the form 
\begin{equation}
\rho\otimes\tilde \rho=\sum_{\tiny{N,M}}\sum_{\tiny{ 
\left.\begin{array}{c}\textbf{n},\textbf{n'} \\ \sum_i n_i=N \\ \sum_i n_i'=N\end{array}\right.}} \sum_{%
\tiny{ \left.\begin{array}{c}\textbf{m},\textbf{m}' \\ \sum_i  m_i=
M \\ \sum_i m_i'= M\end{array}\right.}}\rho_{\mathbf{n},\mathbf{n'}}^{(N)}{%
\tilde \rho}_{\mathbf{m},\mathbf{m'}}^{(M)}|\mathbf{n},%
\mathbf{m} \rangle\langle \mathbf{n}',{\mathbf{m}'} |.
\label{eq10}
\end{equation}
They have the crucial property that for $\sum_in_i\neq\sum_im_i$ one has 
\begin{equation}
\langle \psi_{\mathbf{n},\mathbf{m}}^{+} |\rho\otimes\tilde \rho|\psi_{%
\mathbf{n},\mathbf{m}}^{+} \rangle= \langle \psi_{\mathbf{n},\mathbf{m}}^{-}
|\rho\otimes\tilde \rho|\psi_{\mathbf{n},\mathbf{m}}^{-} \rangle,
\label{eq11}
\end{equation}
as one can easily show using the definition of the eigenstates (see \Sref%
{sec:Proof}) and \Eref{eq10}. This property is used to allow for errors in
the assignments of the measured eigenvalue, that do not alter the average
value. An alternative way of thinking about this is the following. We want
to determine the purity of $\rho$ via the expectation value $\textrm{Tr}%
\{\rho^2\}=\textrm{Tr}\{V_2\rho\otimes \rho\}$. However $V_2$ is not the only
operator whose expectation value in the state $\rho\otimes \rho$ is equal to 
$\textrm{Tr}\{\rho^2\}$. Below we will introduce a whole family of operators $%
V_2^{(f)}$ (depending on some function $f$), which share this property. 
We define 
\begin{equation}
P_{\pm}(\mathbf{n},\mathbf{m})=\frac{1}{2}(|\mathbf{n},\mathbf{m}
\rangle\pm|\mathbf{m},\mathbf{n} \rangle)(\langle \mathbf{n},\mathbf{m}
|\pm\langle \mathbf{m},\mathbf{n} |),  \label{eq12}
\end{equation}
and further $Q(\mathbf{n},\mathbf{m})=\frac{1}{2}(P_{+}(\mathbf{n},%
\mathbf{m})-P_{-}(\mathbf{n},\mathbf{m}))$, such that we can write $%
V_2=\sum_{\mathbf{n},\mathbf{m}}Q(\mathbf{n},\mathbf{m})$. The factor $1/2$
in the definition of $Q$ is to correct for double-counting. We can divide
the eigenstates of $V_2$ in two classes, those with $\sum_i n_i=\sum_i m_i$
and the others, leading to the representation 
\begin{equation}
V_2=\!\!\!\!\!\!\sum_{\tiny{\left.\begin{array}{c}   \textbf{n},\textbf{m}%
\\\sum_in_i=\sum_im_i\end{array}\right.}}\!\!\!\!\!\!Q(\mathbf{n},\mathbf{m}%
)+\!\!\!\!\!\!\sum_{\tiny{\left.\begin{array}{c} \textbf{n},\textbf{m}%
\\\sum_in_i\neq\sum_im_i \end{array}\right.}}\!\!\!\!\!\!Q(\mathbf{n},\mathbf{m}).
\label{eq13}
\end{equation}
Note that due to \Eref{eq11}, the expectation value of each term
in the second sum is identically zero for states of the form \Eref{eq10}.
Therefore, the expectation value of the Operator $V_2$ is the same as the
one of $V_2^{(f)}$ defined as 
\begin{equation}
V_2^{(f)}=\!\!\!\!\!\!\sum_{\tiny{\left.\begin{array}{c}   \textbf{n},\textbf{m}%
\\\sum_in_i=\sum_im_i\end{array}\right.}}\!\!\!\!\!\!Q(\mathbf{n},\mathbf{m}%
)+\!\!\!\!\!\!\sum_{\tiny{\left.\begin{array}{c}  \textbf{n},\textbf{m}%
\\\sum_in_i\neq\sum_im_i\end{array}\right.}}\!\!\!\!\!\!f(\mathbf{n},\mathbf{m})Q(\mathbf{n},%
\mathbf{m}),  \label{eq14}
\end{equation}
where $f(\mathbf{n},\mathbf{m})$ is an arbitrary (real-valued) function. In
particular we can choose 
\begin{equation}  \label{eq15}
f(\mathbf{n},\mathbf{m})=\left\{%
\begin{array}{cc}
0 & s(\mathbf{n}+\mathbf{m})=1 \\ 
1 & s(\mathbf{n})=s(\mathbf{m})=s((\mathbf{n}+\mathbf{m})/2) \\ 
-1 & s(\mathbf{n})=s(\mathbf{m})\neq s((\mathbf{n}+\mathbf{m})/2)%
\end{array}%
\right.,
\end{equation}
where we use the notation $s(\mathbf{x})=\sum_i x_i \textrm{ mod } 2$. The
eigenstates of $V_2^{(f)}$ are the same as those of $V_2$, but they have
different eigenvalues. The advantage of choosing $f$ in the above form is
that the eigenvalue of $V_2^{(f)}$ can be determined form the total number
of atoms and the number of atoms in copy one after the beam splitter, as
presented in \tref{table1}. This can easily be checked by explicitly
looking at the transformation of all the different classes of eigenstates of 
$V_2^{(f)}$ under the beamsplitter. 

\section{Mixtures of Copies}

\label{mixtures_of_copies} As pointed out in \cite{Ekert:2002fp} the
expectation value of the operator $V_2$ in the state $\rho_{\mathrm{tot}%
}=\rho\otimes\rho$ is given by $\textrm{Tr}\{\rho^2\}$. The assumption of having two completely uncorrelated copies of the form $\rho_{\mathrm{tot}%
}=\rho\otimes\rho$, although being difficult to validate \cite{vanEnk:2009eo} experimentally, is a natural one for such systems in optical lattices, where the two copies can be decoupled by a high potential barrier. However, correlated errors, such as for
example global fluctuations in the intensities of the laser beam that generates the
lattice \cite{Pichler:2012hz}, affect the two copies in the same way and lead
to correlations between the copies. Such identical, but correlated errors in
general lead to a state which is a mixture of different products of the same
state, that is $\rho_{\mathrm{tot}}=\sum_i p_i \rho_i\otimes \rho_i$ with $%
\sum_i p_i=1$, $p_i\geq 0$. In the following we show that even in this situation the
measurement of the operator $V_2$ provides useful information about the
entropy in the sense that it provides bounds on the purity of the reduced
system of one of the two ``copies'', that is of the state $\rho\equiv \textrm{%
Tr}_{\mathrm{copy 2}}\{\rho_{\mathrm{tot}}\}=\sum_i p_i \rho_i$. To this end we show that
\begin{equation}  \label{eq17}
\frac{1}{2}\textrm{Tr}\{V_2 \rho_{\mathrm{tot}}\}\leq \textrm{Tr}\{\rho^2\}\leq 
\textrm{Tr}\{V_2 \rho_{\mathrm{tot}}\}.
\end{equation}
We first show the upper bound. Consider the function 
\begin{equation}
F(\{p_i\})\equiv \textrm{Tr}\{\rho^2\}-\textrm{Tr}\{V_2 \rho_{\mathrm{tot}}\} 
=\sum_{i,j}p_i p_j\textrm{Tr}\{\rho_i \rho_j\}-\sum_i p_i\textrm{Tr}\{\rho_i^2\},
\label{eq19}
\end{equation}
defined in the polytope spanned by the $p_i$ with $\sum_i p_i=1$, $p_i\geq 0$. 
Note that this function is zero at the corners of this polytope given by $%
p_i=\delta_{i,j}$. Further, the hessian matrix of $F$ is constant and given
by $H_{i,j}\equiv\frac{d^2 F}{dp_i dp_j}=2 \textrm{Tr}\{\rho_i \rho_j\}$. This
is a Gramian matrix and thus positive semidefinite. Thus the function $F$ is
convex everywhere and its value on the (convex) polytope defined by $\sum_i
p_i=1$, $p_i\geq 0$ is smaller than its value at the corners. This proves the upper
bound. To prove the lower bound we calculate the (unique) global minimum of $%
F$. It is assumed at the position that satisfies $dF/dp_i\equiv\sum_j2p_j%
\textrm{Tr}\{\rho_i\rho_j\}-\textrm{Tr}\{\rho_i^2\}=0$. That is, at the minimum
we find $\sum_jp_j\textrm{Tr}\{\rho_i\rho_j\}=\frac{1}{2}\textrm{Tr}\{\rho_i^2\}$%
. Using this in \Eref{eq19} we find that $F\geq -\frac{1}{2}\sum_i p_i 
\textrm{Tr}\{\rho_i^2\}=-\frac{1}{2}\textrm{Tr}\{V_2\rho_{\mathrm{tot}}\}$, and
thus $\frac{1}{2}\textrm{Tr}\{V_2 \rho_{\mathrm{tot}}\}\leq \textrm{Tr}%
\{\rho^2\} $. These inequalities trivially hold also for purities of
subsystems of one of the two copies (with the corresponding swap-operators). 

The inequalities \Eref{eq17} give bounds in terms of the expectation value
of $V_2$ on arbitrary mixtures of copies. In the typical experiment
situation one expects to be very close to a simple product of two states
with a small admixture of other copies, such that $p_0=1-\epsilon$ with $%
\epsilon=\sum_{i\neq 0}p_i\ll1$. Because the function $F(\{p_i\})$ is convex we can bound 
\begin{eqnarray}
&F(\{p_i\})\geq F(\{\delta_{i,0}\})+\sum_i (p_i-\delta_{i,0})\frac{%
dF(\{p_i\})}{dp_i}\biggr|_{p_i=\delta_{i,0}}   \\
&=-\sum_{i\neq 0} p_i\textrm{Tr}\{(\rho_i-\rho_0)^2\} \geq-2\sum_{i\neq 0}
p_i\equiv -2\epsilon,
\end{eqnarray}
and thus one has $\textrm{Tr}\{V_2 \rho_{\mathrm{tot}}\}-2\epsilon\leq \textrm{Tr%
}\{\rho^2\}\leq \textrm{Tr}\{V_2 \rho_{\mathrm{tot}}\}$.
\newline

\bibliographystyle{iopart-num.bst}
\bibliography{mybiblio,HannesBiblio}

\end{document}